\documentclass[12pt]{article}

\usepackage{graphicx}

\usepackage{enumerate}
\usepackage{amsmath,amssymb}
\usepackage{authblk}
\usepackage{hyperref}
\usepackage{newclude}

\usepackage{float}
\usepackage[multiple]{footmisc}

\title{\LARGE Hints towards the Emergent Nature of Gravity}

 	\author[1]{Niels S. Linnemann\thanks{niels.linnemann@unige.ch}} 
  	\author[2]{Manus R. Visser\thanks{m.r.visser@uva.nl}}

\affil[1]{
\it{Geneva Centre for Philosophy of Science, University of Geneva, 

2 rue de Candolle,  CH-1211 Geneva 4, Switzerland }}
\affil[2]{
\it{Institute for Theoretical Physics,  University of Amsterdam, 

Science Park 904, 1098 XH Amsterdam, The Netherlands}  }

\date{\today}

\usepackage{xargs}                      
\usepackage[pdftex,dvipsnames]{xcolor}  

\usepackage{parallel}

\usepackage[colorinlistoftodos,prependcaption,textsize=tiny]{todonotes}
\newcommandx{\unsure}[2][1=]{\todo[linecolor=red,backgroundcolor=red!25,bordercolor=red,#1]{#2}}
\newcommandx{\change}[2][1=]{\todo[linecolor=blue,backgroundcolor=blue!25,bordercolor=blue,#1]{#2}}
\newcommandx{\info}[2][1=]{\todo[linecolor=OliveGreen,backgroundcolor=OliveGreen!25,bordercolor=OliveGreen,#1]{#2}}
\newcommandx{\improvement}[2][1=]{\todo[linecolor=Plum,backgroundcolor=Plum!25,bordercolor=Plum,#1]{#2}}
\newcommandx{\thiswillnotshow}[2][1=]{\todo[disable,#1]{#2}}

\makeatletter
\newlength{\@testbar}%
\newlength{\@thusbar}%
\newif\ifthusstar
\newcommand*{\thus}{\@ifstar{\thusstartrue\@thus}{\@thus}}%
\newcommand*{\@thus}[1]{%
  \settowidth{\@testbar}{$\therefore$ #1}%
  \setlength{\@thusbar}{\minof{\@testbar}{\linewidth}}%
  \ifhmode\par\fi
  \vspace*{-1ex}\rule{\@thusbar}{0.3pt}\vspace*{-1ex}%
  \item\ifthusstar\else$\therefore$~\fi #1}%
\makeatother

\usepackage{setspace}

\usepackage[round]{natbib}   
\usepackage{braket}

\begin{document}
 
 \onehalfspacing   

\maketitle

\thispagestyle{empty}

\abstract{A possible way out of the conundrum of quantum gravity is the proposal that general relativity (GR) 
 emerges from an underlying microscopic description.   Despite recent interest in the emergent gravity program within the physics as well as the philosophy community, an assessment of the theoretical evidence for this idea is lacking at the moment. We intend to fill this gap in the literature by discussing the main arguments in favour of   the hypothesis that the metric field and its dynamics are emergent. 
First, we distinguish between microstructure inspired from GR, such as through quantization or discretization,  and microstructure that is  not directly motivated from GR, such as strings, quantum bits  or condensed matter fields. The emergent gravity  approach can then be defined as the view that the metric field and its dynamics are derivable from the latter type of microstructure.
Subsequently, we assess in how far the following properties of (semi-classical) GR are suggestive of underlying microstructure: (1)  the metric's universal coupling to matter fields, (2)   perturbative non-renormalizability, (3) black hole thermodynamics, and (4) the holographic principle. 
In the conclusion we formalize the general structure of the plausibility arguments put forward. } \\

\newpage
 
 \setcounter{tocdepth}{2}
\tableofcontents 

\newpage 

\section{Introduction}
Gravity is notoriously harder to quantize than other interactions. It is well-known that combining general relativity (GR) with quantum mechanics poses several difficulties which are standardly traced back to the perturbative non-renormalizability of   general relativity. Proponents of quantum gravity (QG)  approaches proper (such as loop quantum gravity, asymptotic safety  and causal dynamical triangulation) nevertheless expect that gravity is after all non-perturbatively renormalizable, and thereby possibly fundamental.  In contrast, according to other approaches -- like   string theory\footnote{  See \cite{Polchinski:1998rq,Polchinski:1998rr}, \cite{Zwiebach:2004tj} and \cite{Becker:2007zj}  for standard textbooks on string theory.} and emergent gravity accounts by \cite{Sakharov1967}, \cite{Jacobson}, \cite{Padmanabhan1}, \cite{Verlinde}, and \cite{Hu} --  the gravitational field   cannot be fundamental but rather emerges from underlying microstructure. 
These approaches can be subsumed under the term ``emergent gravity''  (EG).

The idea that gravity and spacetime originate from some   underlying microscopic reality in which they do not exist is a widely held belief in, for example,  the string theory community.\footnote{Cf. \cite{WittenReflections},  \cite{Seiberg}, \cite{Verlinde}.} The possibility that gravitational degrees are rooted in non-gravitational degrees of freedom is often supported by reference to the  AdS/CFT correspondence.\footnote{ See \cite{Maldacena} for the original derivation of the AdS/CFT correspondence, and \cite{Aharony} for a comprehensive review.} The AdS/CFT correspondence, however, only provides a straightforward\footnote{ For a more nuanced take on the notions of fundamentality and emergence in the context of AdS/CFT, see \cite{Rickles}, \cite{Teh},  \cite{Dieks}, \cite{Haro} and \cite{Vistarini}.} gateway to a theory of emergent gravity under the  controversial   assumption that the CFT side of the correspondence is more fundamental than the AdS side.
Now, it is precisely the philosopher of physics' task to reflect on such current trends in quantum gravity research. In our view a particularly pressing question in this field is \emph{why} it is widely believed that gravity is emergent -- and \textit{what} is meant by the term ``emergent gravity'' in the first place.

As a matter of fact, it is surprising that the current philosophical literature\footnote{Cf. \cite{Bain}, \cite{Mattingly}, \cite{Knox2013}, \cite{Crowther, Crowther2016}, \cite{Dieks}, \cite{Haro} and \cite{Carlip}. Admittedly, Carlip's article is an appraisal of the emergent gravity program, but it only focuses on the challenges for such a program rather than the positive hints for the emergent nature of gravity.} 
on EG  mainly focuses on notions of emergence or possible implications for QG \textit{rather than appraising general arguments for the emergent nature of GR in the first place}! We intend to fill this gap in the literature by reconstructing and assessing arguments for the high-levelness of gravity, which are often only implicitly contained in the relevant physics literature. 
These arguments  do not assume any specific QG model, but rather consist of pointing out features of gravity which are (at least typically) characteristic of systems with underlying microstructure. In other words, certain features -- ranging from universal coupling to black hole thermodynamics -- allow for mounting arguments from analogy to the effect that the metric field and its dynamics are emergent.  Arguably, these gravitational features even remain ``mysteries'' in a fundamental take on the metric and appear to be in need of explanation,  which however can be easily achieved in an emergent gravity framework.  

The matter is worth exploring in particular in light of the interesting consequences a high-level view of GR could have for the conception of spacetime: it serves promoting a physical view of the metric as one field among others, as opposed to a conventional (geometric) spacetime view as advocated by \cite{Friedman} and \cite{Maudlin}. Hence emergent gravity is very much in line with the particle physics view on GR (the ``spin-2 approach'') advocated by \cite{Feynman} and \cite{Weinberg},  and with the dynamical approach to relativity by \cite{Brown}.
  In any case,
 if one agrees with the call to philosophers by \cite{HuggettWuethrich}  to engage with conceptual problems in the development of theories of QG, an assessment of the   indications for the emergent nature of gravity will count as a sensible, if not necessary, contribution. 
 
  This paper is organised as follows. 
We start out in section \ref{whatiseg} by clarifying our notions of emergence, microscopic structure and high-levelness. Ultimately, we provide a definition of emergent gravity which we take to capture best what is \textit{meant} by emergent gravity in the physics literature. 
In section \ref{indications}, we work through individual features of gravity which can at least naively be taken as suggestive of emergence. 
In section \ref{conclusion} we end by assessing the general structure of the arguments put forward.

\section{What is emergent gravity?}
\label{whatiseg}

\subsection{Definition of emergent gravity}

Emergent gravity is roughly taken to be the view that gravity arises due to the ``collective action of the dynamics of more fundamental non-gravitational degrees of freedom'' (\cite{HuggettWuethrich}). The  aim of this section is to sharpen this rough characterization of emergent gravity by providing answers to the following core questions: (i)  ``emergence of what?'', (ii) ``emergence from what?'' and (iii)  ``what kind of emergence?''. The first question asks what properties of   gravitational physics are actually emergent. The second is concerned with the existence and nature of the underlying microstructure, and the third   focuses on the emergence relation between the gravitational and non-gravitational degrees of freedom. In the next three subsections we   treat these questions separately   noting though that (i) and (ii) are not fully independent.

\subsubsection{Emergent aspects of gravity}

\label{sec:definition}

Which aspects of gravity are emergent? Needless to say the answer to this question depends on what we take gravity to be. A gravitational theory has for instance been  defined by \cite{ThorneLee} as a spacetime theory that reproduces Kepler's laws.\footnote{We thank Dennis Lehmkuhl for referring us to this definition. However, this definition seems to unfairly exclude  modifications of GR (such as higher curvature theories and bimetric theories, like TeVeS)  as gravitational theories.}  In any case, our current best theory of gravity is GR, which formally encompasses a differentiable, 4-dimensional manifold with tensor fields on top. One of the tensor fields necessarily corresponds to a Lorentzian metric field $g$ which has the usual chronogeometric meaning.\footnote{In particular, sufficiently accurate clocks travelling along a wordline segment $\gamma$ will display a reading proportional to the worldline interval $\Delta s= \int_{\gamma} g_{ab}dX^{a} dX^{b}$. Whether the metric intrinsically contains chronogeometric signifiance or gains it dynamically, is up to philosophical debate.
It is, however, not relevant for the characterization of emergent aspects of gravitation.} The metric field satisfies the Einstein equations, $G_{ab} = \frac{8\pi G}{c^4} T_{ab}$, which relate the Einstein tensor $G_{ab}$ to the stress-energy tensor $T_{ab}$ obtained from all other tensor fields (the so-called matter fields). The matter fields themselves are subject to further (possibly coupled) dynamical equations.

One might think that ideally in an emergent model of GR all three aspects emerge together: the manifold, the metric field and its dynamics.\footnote{\cite{LamWuthrich} argue that GR does not necessarily have to be recovered completely in such a scenario: a functionally adequate subset of   notions of GR or empirically adequate modifications of it might be sufficient already.} The emergence of the manifold is, however, somewhat less important, since its physical significance is doubtful. Arguably, the (active) diffeomorphism invariance of fields in GR bereaves the manifold points of any ``thisness" and   renders it as a mathematical scaffolding structure to define physical fields on (cf. the hole argument of \cite{NortonEarman}).  

In some approaches, such as the thermodynamic route to GR by \cite{Jacobson}, the metric field is first of all given ab initio, and then its (Einstein) equations of motion are derived. We think, however, that a genuine account of emergent gravity should  include the emergence of both the metric field and its dynamics (rather than just the dynamics). This is not to say that Jacobson's account is not a form of emergent gravity. In fact, in section \ref{taxonomy} we categorize it as an emergent gravity scenario, because the  thermodynamic interpretation of the dynamical  equations associated to the metric field $g$  naturally leads to the emergent nature of $g$ itself.

In the next subsection, we will further see that  not only the metric field and its dynamics are emergent  aspects of gravity, but    also the  graviton -- the spin-2 particle that mediates the gravitational force -- should count as ``gravitational'' (and  hence  as emergent). According to the spin-2 approach the metric field and its dynamics namely arise through the self-coupling of the graviton,  cf.  \cite{Deser:1969wk}. If we  viewed the spin-2 field as non-gravitational, then GR would emerge from quantized GR, which is however against the intuition behind emergent gravity.

\subsubsection{Two types of microtheories}

\label{sec:microstructures}

In the EG literature,  gravity is unanimously taken to be emergent from (novel) microstructure underlying the metric degrees of freedom. For instance, \cite{Carlip} defines emergent gravity as: \begin{quote}
	``[...] the basic picture is that gravity, and perhaps space or spacetime themselves, are collective manifestations of very different underlying degrees of freedom.'' (p. 200) \end{quote} In a similar spirit, \cite{Sindoni2012} writes: \begin{quote} ``As a provisional definition, we will intend as emergence of  a given theory  as a reorganization of the degrees of freedom of a certain underlying model in a way that leads to a regime in which the relevant degrees of freedom are qualitatively different from the microscopic ones.'' (p. 2)
\end{quote}  

\noindent \cite{Padmanabhan1,Padmanabhan2}, \cite{Hu} and \cite{Verlinde} employ similar notions of EG. They tend to compare (or even equate) the idea of gravity having underlying microstructure (sometimes referred to as ``atoms of spacetime") to the sense in which thermodynamic or hydrodynamic systems have underlying microstructure. 

However, despite the apparent consensus in the literature, this definition of emergent gravity   is problematic.\footnote{Admittedly, some authors like \cite{Sindoni2012} do not even attempt to give a fully worked-out definition.} In general, the \textbf{received definition of emergent gravity} in terms of \textit{very different} underlying degrees of freedom,  or \textit{novel} microstructure simply suffers from vagueness. It is vague both with respect to qualifiers like ``very" or ``novel", and the characterization of what there actually is  supposed to be over and above gravity (``different degrees of freedom" or ``microstructure"). In the following, we give three major instances of changes in degrees of freedom for which it is not a priori clear whether they should be counted as ``very different'' degrees of freedom or not.

Firstly, the discretization of a theory of gravity changes its degrees of freedom in some sense. After all, the continuous version can only be reproduced upon coarse-graining again. But, according to the EG proponent, the degrees of freedom of a discretized theory  should   not be  ``very" or ``qualitatively" different from the degrees of freedom of the original theory. If that were the case, loop quantum gravity (LQG) and causal set theory would   count as instances of emergent gravity, which would make the definition of EG too broad.

Secondly, quantization of a theory does to some extent change the theory's degrees of freedom. As quantization generically leads to discretization of classically continuous quantities -- such as energy, angular momentum, and possibly even spacetime structure --  coarse-graining is typically part of restoring the classical limit (together with suppressing superposition effects). However, quantization does not \emph{necessarily} entail a change of degrees of freedom, at least not in a Bohmian interpretation of quantum mechanics. Whether or not quantization leads to ``qualitatively different'' degrees of freedom  depends on the specific interpretation of quantum mechanics.  Preferably  the definition should not depend upon the interpretation of quantum mechanics at play.

	Thirdly,  in a quantum field theory (QFT) setting,   the renormalization group flow describes a change of theories under a coarse-graining operation. Coarse-graining here corresponds to integrating out high energy   modes.  If one starts  with a renormalizable theory which is well-behaved up to arbitrarily high energies, then coarse-graining will produce effective field theories which are only valid up to a certain level of energy. 
		Despite a change of its degrees of freedom  under renormalization (energy modes are integrated out), the resulting theory will not necessarily involve ``qualitatively" different degrees of freedom. Generally speaking, the   renormalization group flow only changes the coupling constants of a quantum field theory.
		In the context of gravity, asymptotic safety claims that quantum general relativity (quantum GR\footnote{By ``quantum GR'' we denote any form of quantization approach turning GR into a quantum theory, including perturbative quantum gravity, canonical quantum gravity, loop quantum gravity, causal dynamical triangulation, and more; but not, for instance, string theory and causal set theory.}) is subject to a renormalization group flow such that it  eventually runs into a UV-fixed point in theory space. The only difference in the degrees of freedom of     low-energy and high-energy versions of quantum GR is then encoded in the UV cut-off of the corresponding theory.
 
The three examples above show that the received definition of EG is problematic, since it is unclear whether  discretization, quantization, and renormalization lead to ``qualitatively different'' degrees of freedom. This issue is important because we   want to distinguish EG from other quantum gravity accounts   that arise from discretizing, quantizing or renormalizing  GR. One could attempt to sharpen the characterization of the degrees of freedom underlying gravity by calling them  ``non-gravitational" (see also \cite{HuggettWuethrich}), but this remains unsatisfactory as long as one has no clear notion of  ``gravitational".  We already alluded in the previous subsection to the problem that we cannot straightforwardly give a definition of  ``gravitational'' in terms of GR alone, without running the risk of calling theories underlying GR non-gravitational, which are however not at all instances of emergent gravity. For example,   the spin-2 approach, LQG, and causal set theory are possible quantum theories underlying GR that we would like to  exclude as cases of emergent gravity.  

We now propose to circumvent this issue by introducing a distinction between two types of microtheories, i.e. two types of theories underlying a  theory $T$:\footnote{We necessarily require microtheories to $T$ to reduce to $T$ in an appropriate limit, such as a classical or low-energy limit.}

\begin{quote}
We call a theory $M_1$ underlying a    theory $T$  a \textbf{type I microtheory} to $T$ if and only if it is inspired\footnote{We would like to acknowledge Jeremy Butterfield for suggesting this term to us, and  thank Claus Beisbart and Sam Fletcher for pressing us on our definition.}
 from $T$ (for instance through discretization, quantization or renormalization). An underlying theory $M_2$ to theory $T$ is called a \textbf{type II microtheory} if and only if it  is not directly inspired or motivated\footnote{One might argue that  notions as ``inspired'' or ``motivated'' from are   as vague as the notions ``qualitatively'' or ``very different'', criticized above. However,  despite their granted vagueness, the former notions allow for a more precise distinction between what is referred to as EG in the literature and what is not, as we demonstrate below.} from  $T$. 
\end{quote}

\begin{quote}
The structure  linked to a microtheory is called \textbf{microstructure}.
\end{quote}

\begin{quote}
A   theory is  called  \textbf{emergent} if and only if there exists an underlying type II microtheory to it.
\end{quote}

\noindent 
 A type I microtheory to GR can now coherently be counted as gravitational, whereas a type II microtheory to GR is by definition non-gravitational since gravity emerges from it. Another advantage of this \textbf{new definition of emergent gravity} in terms of type II microtheories is that the microstructure associated to these theories is automatically novel, because   it cannot be inspired from gravity. Examples of possible type I microtheories to GR are causal set theory, loop quantum gravity and other quantum gravity  approaches proper.   This is in line with the general idea of emergent gravity being something over and above quantum gravity proper (see \cite{Carlip}, \cite{Padmanabhan1}, \cite{Jacobson}). Furthermore, string theory would count as an instance of type II microtheory to GR.\footnote{Arguably, perturbative string theory counts as a type I  microtheory, as its target space with a dynamic background metric is inspired from GR, whereas  non-perturbative string theory is a type II microtheory underlying GR, since it does not involve any kind of Lorentzian metric.}

 One might object to our definition of emergent gravity that  a quantum GR approach can already lead to   a quite different picture of spacetime. For instance, many solutions in LQG will not in any sense reduce to general relativistic spacetimes. So our  distinction of quantum gravity approaches based on whether or not ``they are inspired from GR" seems to be independent of how radically relativistic spacetimes emerge from underlying microstructure. 
Our concern is, however, not with radicalness as we do not and cannot claim that quantum GR approaches do not provide spectacular instances of emergence. 
But rather than distinguishing approaches based on the radicalness of emergence at play, our proposed definition of EG is designed to provide a demarcation line between what is generally understood as the emergence of gravity and other accounts (such as loop quantum gravity and causal set theory). With this definition at hand, we are then able to  evaluate  the plausibility arguments put forward in section \ref{indications}, in particular in how far they are able to establish the emergence of gravity. The concrete criterion of whether or not an account has been inspired from GR  seems to fulfill this practical purpose best.


As a final note  we  emphasize that,   although proper quantum gravity and emergent gravity approaches are   defined in a different way (as originating from type I and type II microtheories, respectively) they are not necessarily in conflict with one another.\footnote{An often expressed sentiment by proponents of emergent gravity is that gravity should not be quantized if gravity is emergent (cf. \cite{Jacobson}, \cite{Padmanabhan1}, \cite{Verlinde}). One can, however, \textit{prima facie} equally quantize high-level as well as low-level theories. Whether to quantize or not to quantize, is a question of the theory's formal nature, not of the level of reality we apply it to. For instance,   quantization of condensed matter field theories is unproblematic and is well-known to yield quasi-particles, like phonons.} They are trivially compatible as any physical theory (and so even the most successful theory of quantum gravity proper) could in principle  turn out to be underlied by a more fundamental theory.\footnote{This skepticism is quite conventional within a   quantum field theory setting. In lack of knowledge of possible higher energy modes at play in a perturbatively considered QFT,    a  naively fundamental (i.e. renormalizable) QFT  might equally be just effective (see \cite{CaoSchweber}).}
This general skepticism becomes most interesting if it is connected to concrete approaches which are standardly believed to be complete, but are then shown to have underlying structure nevertheless.  Quantized gravity approaches like loop quantum gravity or causal dynamical triangulation (instances of QG) could in fact be underlied by string theory (an instance of EG) \textit{even if} they are UV-complete.
An appraisal of emergent gravity should take this general skeptic's point into account, if it does not want emergent gravity to sound like a cheap option always available.

\subsubsection{Different notions of emergence}

 With respect to the third question --  ``what kind of emergence?'' --  it is important to point out  that philosophers and physicists use the word ``emergent'' differently, and themselves not at all coherently either.\footnote{We would like to thank Karen Crowther  for sharing an unpublished note, in which she distinguished between the philosopher's, the physicist's and the philosopher of physics' usage of the word ``emergence''. This distinction goes back to \cite{Nickles}.}  \emph{Philosophers} tend to  define emergence as the failure of  reduction. A theory (property) is then called emergent from another  more fundamental  theory (property) if the former is not  derivable or explainable from the latter.

In contrast, \emph{philosophers of physics} mostly work with Butterfield's notion of emergence in terms of supervenience on the one hand, and novelty and robustness on the other hand.\footnote{ This understanding of emergence in terms of dependence (supervenience) and independence (novelty or autonomy) is arguably not the only one in the current philosophy of physics' literature -- albeit that it is very dominant. One might for instance also consider pure epistemic emergence in light of unpredictability of one level from another. Also see  \cite{Rickles},   \cite{Crowther2016}, \cite{Vistarini} and \cite{deHaronew} -- among others -- for more recent work.}
 Supervenience of a theory (property) on another (standardly more fundamental) theory (property) means that there can be no change in the former theory (property) without a change in the latter theory (property). The problem with this notion of emergence, of course, lies in its deference of the difficulty in defining emergence to that of defining robustness and novelty (similar to how the philosopher's sense of emergence seems to ultimately depend on the definition of reducibility). Following \cite{ButterfieldEmergence}, we take ``novel'' properties to mean  features which are not contained in the microstructure, whereas ``robust" properties are properties that are not sensitive to (and somehow independent from) the specifics of the underlying structure.

Lastly, \emph{physicists} often use the word ``emergence'' in the sense of reducibility, that is derivability or explainability of a certain structure from another one. Take a string theorist claiming that he/she derived general relativity from perturbative string theory in a low-energy limit: he/she will probably call this an instance of emergent gravity. More concretely, \cite{Seiberg}   takes space and time to be emergent if they ``will not be present in the fundamental formulation of the theory and will appear as approximate semiclassical notions  in the macroscopic world" (p. 1). Here, Seiberg identifies emergence of spacetime with its derivability (through semiclassical approximation) from a microscopic structure.

Which notion of emergence is applicable to emergent gravity? We should not forget that the notion of emergent gravity is  introduced by physicists -- not by philosophers. It is thus plainly wrong to presume that the notion of emergence in EG should be interpreted in  the philosopher's sense. Rather, it is   the physicist's notion which seems to apply here -- as physicists  expect GR to be reducible to more fundamental microscopic theories -- thereby excluding the philosopher's notion from the start.

Still, we do not take the physicist's reduction to exhaustively specify the relationship between the metric  and its  (putative) underlying microstructure. There is arguably an aspect of novelty and robustness in the emerging metric structure which need to be accounted for.\footnote{Some physicists might also   have this notion of emergence in mind. The previous distinction between the three senses of emergence should  be understood as a general labeling, not specific to the context of QG. We thank Christian W\"{u}thrich for pressing us on this point.}
 First, the non-gravitational degrees of freedom (as defined in the previous section) are qualitatively different  from (and hence \emph{novel} with respect to) the gravitational degrees of freedom, as the former are not inspired from the latter. Secondly, albeit more specifically, in thermodynamic approaches by Jacobson, Verlinde and Padmanabhan, gravity is viewed as \emph{robust}  with respect to the underlying microstructure, in the sense that it only depends on   coarse-grained notions such as entropy and temperature. Thus, the notion of emergence in emergent gravity  seems most suitably   characterized by the philosopher  of physics' sense. This also seems to capture best how advocates of EG models use the word ``emergence''.

\subsection{Relation to unification, emergent spacetime and EFT}

In the previous section we tried to get hold of the notion of emergent gravity -- as used in practice -- by characterizing it through a microtheory which is underlying GR but not inspired from it (type II microtheory). We will now compare the notion of emergent gravity   to related ones.

\begin{description}
\item[Unification.]
The quest for the unification of gravity with the other fundamental interactions of Nature is logically compatible with emergent gravity, as defined above. If gravity is emergent, and unification is realized at a certain energy scale, then also the other interactions  might have type II microstructure. In a sense the idea of   emergent gravity  then  loses its pre-eminent status.  For example, in the   unification scenario of string theory all interactions are on a par with each other and are equally emergent from the underlying microstructure of  strings and D-branes. In other emergent gravity approaches such as those of \cite{Verlinde, Verlinde2016}  and \cite{Jacobson, Jacobson2015}, typically only gravity is viewed as emergent whereas the other forces are treated as fundamental. This, however, does not exclude the scenario that all forces eventually emerge from a joint microstructure. But the approach to such a unified description could be   different from  other  unification scenarios such as string theory, in the sense that gravity is expected to emerge prior and in a qualitatively different way from the other forces when going up the energy scale.

\item[Emergent spacetime.]
In the current GR era ``gravity'' is often used synonymously with ``spacetime'' so that, at first sight, ``emergent spacetime'' and ``emergent gravity''  seem to be synonyms. The usage of the term ``emergent spacetime'', however, comprises many different senses in which typical aspects of spacetime -- such as dimensionality, causal and chronogeometric structure, continuity of space and time, and more -- turn out to be non-fundamental. For example, \cite{HuggettWuethrich} use the term for both the emergence of classical spacetime from their quantum versions, as in LQG,  \textit{and} the emergence of spacetime from another spatiotemporal structure, as in condensed matter approaches,  or non-spatiotemporal structure, as in non-commutative geometry. On our account the former type of emergence is not an instance of EG, whereas the latter is.\footnote{Emergence of spacetime does not  necessarily entail the emergence of matter fields. Only without a straightforward rule   to assign the currently known matter fields to the underlying degrees of freedom of spacetime, the emergence of spacetime implies the emergent nature of  matter as well. 
}

\item[Effective field theory.]
Within a particle physics or quantum field theory approach to GR  the Einstein-Hilbert action is standardly rendered as a low-energy effective action (cf. \cite{Burgess2003}). Similarly, perturbative quantum GR is then simply understood as a low-energy effective field theory (EFT) (cf. \cite{Donoghue}). The EFT view and emergent gravity only make contact if the gravitational low energies degrees of freedom are qualitatively different from the high-energy degrees of freedoms (for example if the spin-2 particle is a composite particle). Mere effectiveness of a field theory, however, does not yet entail that its higher energetic versions feature qualitatively different degrees of freedom. For example, in the asymptotic safety approach to gravity the gravitational action changes with the energy scale while the metric degrees of freedom remain fundamental.
\end{description}

\subsection{Taxonomy of accounts: top-down vs. bottom-up}
\label{taxonomy}

In the previous sections we have given a broad definition of emergent gravity and demarcated this notion from related ones. In this section we give a quick overview of the several models of emergent gravity and different approaches towards finding such models. 

One could taxonomize the emergent gravity program in (at least) two natural ways: either by qualitatively distinguishing different models from each other, or by differentiating between research strategies for establishing the emergent nature of gravity. Regarding the first taxonomy,  \cite{Bain2013} and \cite{Carlip} have classified different models in terms of the nature of the underlying microstructure. In some models the underlying structure is spatiotemporal, whereas in others any sense of ordinary spacetime itself is emergent and hence the microstructure is necessarily non-spatiotemporal. 

In this paper we are more concerned with the second kind of taxonomy. We distinguish between two different strategies towards emergent gravity: (1) bottom-up accounts that start from the underlying microstructure and work their way upwards towards GR; and (2) top-down accounts that start from  the emergent gravitational structure and try to find   universal features of the underlying microstructure.\footnote{`Bottom' (`top') here refers  to UV (IR) physics,  which is opposite the convention of \cite{Hu}.} 
Since finding a detailed microscopic model of emergent gravity is generally quite hard, some authors just study analogies between gravitational features of GR (or QFT in curved spacetime) and standardly emergent structures (such as thermodynamics or hydrodynamics). Such top-down accounts do not establish gravity to be emergent, but should often be taken as   hints towards the emergence of gravity. Conversely, bottom-up accounts typically start from microscopic theories  such as condensed matter physics, quantum information theory or string theory, and try to derive GR. 

Below  we list   different top-down and bottom-up accounts (see \cite{Carlip} for more EG accounts):

\begin{description}
\item[Bottom-up accounts.]  The archetypical example  is induced gravity by \cite{Sakharov1967}, and variants thereof (see \cite{Visser2002} for a review), in which the Einstein-Hilbert action appears as a term in the one-loop effective action of matter fields. Other examples include: non-perturbative string theory, such as  the   matrix model for M-theory by \cite{Banks1997}; some version of the AdS/CFT correspondence, where the conformal field theory is viewed as fundamental and the bulk spacetime as emergent (cf. \cite{Berenstein2005}, \cite{Horowitz}); and possibly analog models of gravity, such as acoustics in a moving fluid (cf. \cite{AnalogueUnruh}, \cite{AnalogueGravity}).

\item[Top-down accounts.]  Typical examples in this category are derivations of the Einstein equations from thermodynamics or hydrodynamics,   such as by  \cite{Jacobson},  \cite{Padmanabhan1} and \cite{Verlinde}. Recently, \cite{Jacobson2015} proposed a new derivation of the Einstein equations  from the entanglement properties of the vacuum. Another approach by \cite{Hu} is to extend the semi-classical field equations by adding fluctuation terms on the energy/matter side.
\end{description}

\section{Hints towards an emergent nature}
\label{indications}

In this section we will construct and criticize arguments to the effect of an emergent gravity paradigm.
Naively speaking, there are two strategies for arguing that gravity is emergent: (1) to look for indications in our best theories of gravity (GR, its higher derivative extensions or semi-classical gravity) for an emergent paradigm; and (2) to directly present a   model for the supposed microstructure from which gravity emerges.

The first strategy is to look for analogies between (X) gravitational theories and (Y) other theories said to have microstructure, such as thermodynamics or hydrodynamics. One then hopes to demonstrate the emergent nature of gravity  by establishing a sufficiently strong analogy between (X) and (Y) in a relevant sense.
Certain features of gravity -- mainly its universality, perturbative non-renormalizability, its thermodynamic features and  its holographic nature -- allow for mounting such arguments from analogy.  Arguably, these gravitational features even remain ``mysteries'' in a fundamental take on the metric and appear to be in need of explanation,  which however  can be easily achieved in an emergent gravity framework. We call this the \emph{top-down strategy} (similarly to the already mentioned top-down accounts).

But how could a top-down strategy ever do more than establishing the plausibility of an emergent nature? Strictly speaking, it cannot do more. But philosophy of physics, among other things,   has to make sense of the conceptual aspects of a theory, and read the signals that come from these conceptual aspects. If there are thus signals calling for an emergent reading of gravity, then the philosopher of physics has to make them public. The real question  is,  rather, how strong these signals are, i.e. how to construct arguments from them and consequently assess their relative strengths.

The second strategy at worst only gives a proof-of-concept for the emergence of gravity. At best, it provides  a genuine candidate for the microstructure underlying gravity. String/M-theory as well as condensed matter systems from which gravity emerges are  examples of such candidates. We call this the \emph{bottom-up strategy} (similarly to the above mentioned bottom-up accounts).

In the current paper we   only follow the top-down strategy. Below we   analyze four hints for the emergent nature of gravity: the universality of  gravity (subsection \ref{universality}), the perturbative non-renormalizability (subsection \ref{nonrenormalizability}), black hole thermodynamics (subsection \ref{spacetimethermodynamics}), and the holographic principle (subsection \ref{holography}). We will see that the latter three hints are closely related.

\subsection{Universality of gravity}
\label{universality}

\label{sec:universalcoupling}

Universal attraction is, of course, a well-known property of gravity, but first we want to clarify what  exactly is meant by this type of universality, both in Newtonian gravity and in general relativity. We will  then  be sufficiently equipped to assess whether the universal nature of gravity hints at its emergence.

In Newtonian gravity, one can distinguish between two aspects of the universality of the gravitational force: its universal scope  and its universal strength. The first universal aspect is that all particles have a gravitational mass, and hence attract each other as described by Newton's law of gravitation
\begin{equation*} 
F = G \, \frac{  m_1 m_2}{r^2} \, . 
\end{equation*}
\noindent Hence, Newton's law of gravitation is \textit{universal in scope} as it applies to any two Newtonian objects of a given mass.  The other fundamental forces (electromagnetic, weak and strong force) do not have such a general scope as gravity. The electromagnetic force, for example, only applies to particles with electric charge. A related important difference between gravitation and electromagnetism is that there exists only one type of gravitational charge, whereas there are two types of electric charges in nature.

The second universal feature of Newtonian gravity is \textit{its universal strength}: gravity is `equally strong' for different types of freely falling bodies. More precisely,  two test bodies in a gravitational field which are not acted upon by other forces, such as air resistance, move with the same acceleration  irrespective of their mass and internal composition.  In the literature this property is known as the `weak equivalence principle' (WEP) or as the `universality of free fall', and it holds as a consequence of the proportionality between inertial and gravitational mass.\footnote{Cf. \cite{Clifford}.} 
Within the ontology of Newtonian theory, WEP is an extra assumption on top of the three laws of Newtonian mechanics, which is based on empirical evidence  (in contrast to GR, where it is built into the formalism). Taken together, the universal scope of gravity and its universal strength imply that \emph{all} test bodies in a gravitational field fall equally fast in the absence of other forces. 

In general relativity, or any sound relativistic theory of gravity, both of these universal  features are of course present, albeit in a different guise. First,   the universal scope is extended to everything that carries energy: the gravitational field in GR acts on \emph{all} sources of stress-energy.\footnote{In GR (or any other metric theory of gravity) the metric field does not only act on sources of stress-energy, but it is also acted upon by these sources. Hence, the metric is a dynamical field in such theories of gravity.} Second, the WEP is also an indisputable principle in general relativity, and it even forms one of the foundations of the theory. One of the main differences, though, between Newtonian gravity and general relativity is that freely falling bodies in GR are moving inertially with respect to a local inertial frame, and hence no  gravitational force  is acting on them (in contrast to the Newtonian picture of gravity as a force). In fact, the  trajectories of  freely falling test bodies in GR are identified with   timelike geodesics of the metric. Hence, the WEP does not follow in GR  from the  equality between the inertial and gravitational mass -- since these are not fundamental notions in GR -- but it is rather embodied by the fact that geodesic motion is independent of the particular constitution of test bodies.\footnote{Following \cite{BrownRead}.}

Furthermore, in standard treatments of GR  the term `universality' is often attributed to the coupling of the metric to matter fields. \cite{Clifford}, for example, defines the universal nature of this coupling as follows:\footnote{This     definition of universal coupling dates back to work by \cite{ThorneLee}. They have also proven that a Lagrangian-based relativistic theory of gravity is universally coupled if and only if it   is a metric theory. 
}
 
 \begin{quote}
  ``The property that all non-gravitational fields should couple in the same manner to a single gravitational field is sometimes called `universal coupling'.''   (p. 68)
\end{quote}

\noindent In this definition both of the universal features mentioned above are apparent: the coupling has a universal scope since it applies to \emph{all} non-gravitational fields, and it has   universal strength as it always occurs \emph{in the same manner}. Admittedly, the   characterization of universal strength is rather vague, but what is probably meant here by ``the same manner'' is that the coupling between matter fields and the one gravitational field is always    \emph{minimal}   (which excludes curvature-matter couplings in the equations of motion of the matter fields). It should be noted, though, that the notion of universal coupling  actually  goes beyond the two universal features considered above. Universal coupling is namely a central aspect of  the Einstein equivalence principle (EEP), which is a stronger claim than the weak equivalence principle.\footnote{It is conjectured, though, by Leonard Schiff that any complete and self-consistent gravitation theory that obeys WEP must also  obey EEP.  See \cite{Clifford}  for an overview of the plausibility arguments in favour of Schiff's conjecture.} The EEP states that in any local Lorentz frames  the non-gravitational laws of physics must take on their familiar special relativistic forms  (cf. \cite{Gravitation}).  This principle   ensures 
that the outcome of any non-gravitational experiment is independent of the spacetime location, and is hence more powerful than the WEP.\footnote{As pointed out by \cite{ReadBrownLehmkuhl}, the EEP in its usual form -- contra common belief -- does not hold strictly. The standard Maxwell wave equations in curved spacetime, for instance, do feature curvature couplings, and thus provide an example of equations of motion in the general relativistic context which -- even at a point -- do not take their special relativistic form. As a result,   the EEP is actually only realized effectively in GR,  namely in the case that local curvature effects are negligible.}

Now, the universality of gravity has been put forward by some authors as a hint towards the emergence of gravity. We will analyze their arguments, and specifically pay attention to which aspect of universality (universal scope or strength) is being used in the arguments. In our analysis we take   `universal coupling'  to be the most precise definition of   universality of gravity in GR. 

To begin with, \cite{Verlinde} argues for emergent gravity as follows in his introduction:

\begin{quote}
``Of all forces of Nature gravity is clearly the most universal. Gravity influences and is influenced by everything that carries an energy, and is intimately connected with the structure of space-time. The universal nature of gravity is also demonstrated by the fact that its basic equations closely resemble the laws of thermodynamics and hydrodynamics. So far, there has not been a clear explanation for this resemblance.'' (p. 1)
\end{quote}

\noindent Admittedly, Verlinde at first sight   seems to conflate (or at least directly equate) two different notions of universality, namely the universal coupling of the gravitational interaction and the universal applicability of thermodynamics and hydrodynamics (also to GR). However, we take him to promote the view that universal coupling of gravity, on the one hand, and the applicability of thermodynamics to GR, on the other hand, find a common cause explanation in some kind of coarse-graining process. He subsequently points out that these universal aspects of gravity would be explained by the emergence of gravity.

The idea that the universal attraction of gravity
 hints at underlying (non-trivial) microstructure already dates back (at least) to Feynman. In section 1.5 of the Lectures on Gravitation \cite{Feynman}  suggested the following:
  
\begin{quote}
``The fact of a universal attraction might remind us of the situation in molecular physics; we know that all molecules attract one another by a force which at long distances goes like   $1/r^6$. This we understand in terms of dipole moments which are induced by fluctuations in the charge distributions of molecules. That this is universal is well known from the fact that all substances may be made to condense by cooling them sufficiently. Well, one possibility is that gravitation may be some attraction due to similar fluctuations in something, we do not know just what, perhaps having to do with charge.'' (p. 15)
\end{quote}

\noindent Feynman raises the interesting possibility that gravity might be a high-level interaction, or in his words ``that gravitation is a consequence of something that we already know, but that we have not calculated correctly'' (p. 15). To make this point, he suggests that the universal coupling property of gravity might equally be a result of coarse-graining as the dipole expansion for a composition of molecules. \textit{All} molecules  attract each other with the same force -- given by the $1/r^6$-dipole expansion -- when   seen from a sufficiently large distance. Since the same applies to gravity -- all particles attract each other with the same $1/r^2$-force -- one could think that this is also the result of a multipole expansion, induced by fluctuations in some underlying microstructure. Finally, note that Feynman uses both aspects of universality in his argument: the universal scope (\emph{all} molecules attract each other) and  the universal strength    (the force between them is always \emph{the same}).

Now, based on  these ideas of Verlinde and Feynman we can construct the following argument from universality towards EG. We will use the similarity between hydrodynamics and gravitation to make the argument more concrete. First of all,   hydrodynamics and gravity   are both universal in scope and in strength. Also in hydrodynamics of uncharged fluids the motion of \emph{all} test bodies in a fluid is in a sense \emph{the same}, i.e. their trajectory does not depend on their internal structure or constitution.
 In hydrodynamics, the universal scope and strength   are  the result of  the  coarse-graining of the underlying microstructure. Different microstructures (such as different types of atoms) lead to the same smooth hydrodynamic description.
 In a similar fashion,  the universal coupling of gravity could also be the result of coarse-graining.   This  argument for EG  can be formalized as follows:

\begin{enumerate}
\item Universality in hydrodynamics implies coarse-graining.
\item Coarse graining implies type II microstructure. 
\item Universal coupling of gravity is similar to universality in hydrodynamics.
\thus Universal coupling of gravity implies type II microstructure.
\end{enumerate}
\noindent

\noindent We take premise (1) to be largely uncontroversial. Premise (2) is, however, easy to attack as renormalization group theory provides a straightforward counterexample to it. Coarse-graining could also follow from the renormalization group flow, and hence be due to type I microstructure (as discussed in section \ref{sec:microstructures}).
Premise (3) can be attacked from a geometrical view\footnote{See for example \cite{Maudlin}.} of the metric field:   the metric field  in some sense plays the role of spacetime and, therefore, the universal coupling between the metric field and matter fields is to be expected. According to this view, the universality of gravity is a natural consequence of the fact that it relates to the general framework of spacetime, and is hence not similar to the universality in hydrodynamics.
However, this specific criticism on premise (3) is moot if the geometric view of the metric field in GR is moot. A rebuttal of the geometric view can, for instance, be based on the  dynamical approach to relativity by \cite{Brown}.
 
 Thus, we conclude that the universal coupling property of gravity is with no doubt striking. But whether it can be taken as a valuable hint for the high-level nature of gravity is far from clear.

\subsection{Perturbative non-renormalizability}
\label{nonrenormalizability}

\label{sec:renormalizability}

A generic quantum field theory contains divergences. In a perturbative series expansion of a QFT, these divergences manifest themselves in loop terms, which involve an integration over arbitrarily high momenta. As a sort of technical cure,   these divergences can be absorbed into a proper redefinition of the coupling constants of the QFT. A theory is commonly called \emph{renormalizable} if only a finite number of coupling constants need to be redefined to remove the mentioned divergences.\footnote{The definition of renormalizability is independent of the perturbative/non-perturbative distinction. In particular, a perturbatively renormalizable QFT is generically still divergent in the sense that the perturbative series is a divergent but asymptotic series. Note though that a perturbative series may still define a QFT uniquely.}

A still influential tradition in quantum field theory, dating back to Bethe,  regards only renormalizable quantum field theories as good candidates for fundamental theories (cf. \cite{CaoConceptualDevelopments}, section 8.3).\footnote{The renormalizability of a theory alone cannot completely establish its fundamentality. This follows from the (physical) possibility that there exist higher, but strongly suppressed, energy modes which are not captured by the QFT under consideration. Hence, it might be the case that the QFT is the result of coarse-graining from yet another higher energetic QFT. In this respect, \cite{CaoSchweber} rightly advocate  (1) an epistemological anti-foundationalism and (2) a plurality in QFT ontology.} Such a view builds on the presuppositions that fundamental theories should be predictive, and that only renormalizable theories can be predictive in arbitrary energy regimes. The latter presupposition is based on the strict definition of predictivity that a theory  possesses predictive power if it contains a finite number of parameters -- which makes (non-)renormalizable theories, by definition, (non-)predictive. 

This traditional view   has important consequences for the (non)-fundamentality of  quantum GR. It is generally believed that GR, viewed as a quantum field theory, is perturbatively non-renormalizable. This is often presented as a consequence of the negative mass dimension of Newton's constant (gravity's coupling constant).\footnote{The mass dimension of Newton's constant is $[G]= {2-d}$, where $d$ is the number of spacetime dimensions, which means that a gravitational theory is power counting non-renormalizable for   $d>2$.} Although the power-counting argument is generically a good criterion to judge whether a theory is renormalizable or not; it is rather heuristic, and neither a necessary nor a sufficient criterion for renormalizability (see for instance \cite{PeskinSchroeder}, section 10.1). More precisely, quantum GR has been established to be non-renormalizable,   up to 2-loop order (\cite{GoroffSagnotti}). It is only an expectation, albeit not a theorem, that quantum GR is also divergent at all  orders in a perturbative expansion.

Thus, according to  the traditional view, GR's (expected) non-renormalizability makes it a non-fundamental theory and thus hints at an emergent nature of the theory. The argument can be formalized as follows:

\begin{enumerate}
\item Non-renormalizability implies non-predictivity.
\item Non-predictivity implies non-fundamentality. 
\item (Quantum) GR is non-renormalizable.
\thus (Quantum) GR is non-fundamental.
\end{enumerate}

\noindent Let us now assess the validity of the three assumptions that feature in this argument. Premise (1) is true  by definition. We take this definition of (non-)predictivity to be rather uncontroversial, with the caveat, of course, that GR can be predictive at a certain low-energy scale (say in the solar system) but non-predictive at the Planck scale. Premise (2) only seems acceptable if one regards physical theories to be solely constructed for predictive purposes, and nothing more.  
Only then can we straightforwardly categorize a theory which is not predictive to arbitrary high energies as   non-fundamental. Standard accounts of physical theories, though, also value other explanatory features than mere predictivity, such as their ability to explain structures of the world. 
Therefore, a more generous -- in other words, less narrow -- account of theories than that of Bethe can simply accept that a fundamental theory of the world turns out to be unpredictive in certain regimes, as long as it sufficiently satisfactory in other aspects. 

Whereas premise (2) is subject to philosophical controversy, premise (3) is simply an open question within the current state of physics. Quantum GR has so far only been convincingly established to be perturbatively non-renormalizable, and there exist after all paradigm cases of theories which are perturbatively non-renormalizable, but non-perturbatively renormalizable, such as the \cite{Parisi} model. Whereas many quantizers of gravity like to stress this possibility of the non-perturbative  renormalizability,  in particular string theorists nevertheless keep up the analogy between non-renormalizable gravity and non-renormalizable Fermi 4-theory --  which was replaced by a renormalizable higher energy theory --  in   demand for new, underlying structure to GR.\footnote{In this tradition, \cite{Zee} for instance pathetically asks: ``Fermi's theory cried out, and the new physics turned out to be the electroweak theory. Einstein's theory is now crying out. Will the new physics turn out to be string theory?" (p. 157)}

Arguments for the non-perturbative renormalizability of gravity have been put forward building on holography in one sense or another. The entropy argument by \cite{Shomer} -- originally going back to \cite{Banks} --  for instance tries to establish that $d$-dimensional GR cannot be turned into a $d$-dimensional renormalizable quantum field theory through cashing out an incompatibility of the Bekenstein-Hawking entropy formula with a generic entropy formula for a renormalizable theory. 
However, arguably the semi-classical presuppositions behind this reasoning,  such as the  violent extrapolation of the validity of the Bekenstein-Hawking formula to high energies,  render the argument problematic.\footnote{Cf. \cite{DoboszewskiLinnemann}.}

Going beyond a mere stance on whether GR is renormalizable or not, \cite{Verlinde}  even claims that today's standard framework for renormalization, the Wilsonian framework, is inapplicable to gravity. He points out that   (1) Wilsonian renormalization rests on the insensitivity of   large scale physics from low scale physics, while (2) GR \emph{is} sensitive to  short distance physics as it ``knows'' about the (number of) fundamental degrees of freedom underlying gravity. The latter statement is based on holographic considerations, which imply that there is a limited amount of underlying degrees of freedom associated to each region of space. It is in this sense that gravity knows about the more fundamental level.

\subsection{Black hole thermodynamics}
\label{spacetimethermodynamics}

Thermodynamic phenomena are standardly seen to be underlied by non-trivial microstructure.\footnote{Whether thermodynamics is reducible -- on the level of theories -- to statistical mechanics is a slightly different and highly intricate question. This is    a long-lasting debate in the philosophy community.} For instance,     temperature is viewed as a measure of the averaged motion of underlying particles, and in the Boltzmannian view  entropy is a measure of the amount of underlying microstates for a given macroscopic system.
In this section we will explore in how far the   thermodynamic properties of   black holes (and those of causal horizons in general)\footnote{Cf. \cite{JacobsonHorizon}.} can,  as a consequence, be taken to hint at underlying microstructure of the gravitational field. The   structure of the argument can be formalized as follows:

 \begin{enumerate}
\item Thermodynamic systems have underlying type II microstructure.
\item The laws of black hole mechanics are similar to those of thermodynamics.
\thus Black hole spacetimes have underlying type II microstructure.
\end{enumerate}

\noindent Note that not only the relationship between thermodynamics proper and black hole thermodynamics, but also the link between thermodynamics proper and type II microstructure is far from uncontroversial, as we will point out in the following. However before we do so, we  mention   a stronger version of the analogical argument, which is not based on black hole thermodynamics but on the thermodynamic properties of local Rindler causal horizons through each spacetime point.  In the latter setup,  \cite{Jacobson} has derived the field equations in GR from a thermodynamic first law that governs these local horizons, and thereby he interpreted the Einstein equation as an equation of state. This seminal derivation suggests the following analogical argument  (as also hinted at by Jacobson himself):

\begin{enumerate}
\item Thermodynamic systems have underlying type II microstructure.
\item At each point in spacetime the Einstein  equation can be interpreted as a thermodynamic equation of state.
\thus The Einstein equation is due to underlying type II microstructure of GR.
\end{enumerate}

\noindent Although this argument is more far-reaching than the first argument above, in the sense that it applies to the full Einstein equation instead of a specific solution to it, the thermodynamics of local causal horizons is also less well established than that of black holes. Therefore, in the rest of this section  we will mostly focus  on black hole thermodynamics.

\noindent Black hole thermodynamics consists of a group of laws, originally derived in general relativity  by \cite{BardeenCarterHawking}, which are analogous to (variants of) the standard   laws of thermodynamics (see Table \ref{tab:Kiefer}). Upon results from semi-classical calculations  -- especially the derivation of black hole temperature  by \cite{Hawking} --  the temperature and entropy   featuring in the laws of black hole mechanics were accepted as proper thermodynamic quantities. As a consequence, black hole thermodynamics as such was recognized by the majority of the physics community as   thermodynamics proper.\footnote{See the textbook by \cite{KieferTextbook} or review by \cite{Wald2001}.}

\begin{table}[t!]
\centering
\caption{Overview of the analogies between standard and black hole laws of thermodynamics  
(largely taken from \cite{KieferTextbook}, p. 202)}
\label{my-label}
\vspace{.3cm}
\scalebox{0.8}{
\begin{tabular}{lllll}
 & Standard thermodynamics & Black hole thermodynamics  \\
\hline
Zeroth law:  & $T$ constant on a body in thermal equilibrium  & $\kappa$ constant on the horizon of a black hole    \\
First law:  & $dE=TdS-pdV+\mu dN$ & $dM=\frac{\kappa}{8 \pi G} dA+\Omega_H dJ+\Phi_H dQ$    \\
Second law: & $dS \geq 0$ & $dA \geq 0$   \\
Third law:  & $T=0$ cannot be reached  & $\kappa=0$ cannot be reached   
\end{tabular}
}
\label{tab:Kiefer}
\end{table}

  Contra the received view, \cite{DoughertyCallender} identify two potential problems with black hole thermodynamics. First, they argue that the putative thermodynamic laws of black holes only represent impoverished versions of the actual thermodynamic laws. 
Second, they rightly point out  that the information-theoretic interpretation of black hole entropy   stands in conflict with the  standard non-epistemic interpretation of entropy in thermodynamics. According to a non-epistemic   view on thermodynamics, the description of thermodynamic processes, such as heat cycles in engines, should not be subject to an agent's knowledge. However, proponents of black hole thermodynamics, such as Bekenstein or Wald,  largely maintain a purely agent-based, information-theoretic notion of entropy. \cite{BekensteinOriginal}, for example, justifies the attribution of entropy to the black hole horizon  by stating that otherwise  information  is lost for an agent when something has fallen through the event horizon. 
There are, of course, ways to attack such an information-theoretic reasoning towards entropy, and an agent-based view on thermodynamics more generally (see   also \cite{WuthrichInformation} (2017) for an appraisal of Bekenstein's information theoretic argument).

Rather than taking a general stance in this debate, we are only interested in whether it would have any effect on the analogical argument mentioned above. It seems to us that, generally speaking, the analogical reasoning can still go through even if not all black holes  laws are   proper thermodynamic laws.
What seems most relevant to us is that the analogy holds sufficiently with respect to the notions of entropy implemented in black hole thermodynamics and thermodynamics proper.
Hence, for the evaluation of the analogical argument the decisive questions are   (1) whether  black hole entropy  is a proper thermodynamic entropy,
and   (2) whether thermodynamic entropy is    linked to underlying type II microstructure.

Let us first consider question (1). Against \cite{DoughertyCallender}, recently \cite{Prunkl}, \cite{Curiel} and  \cite{WallaceThermodynamics} have argued for a standard thermodynamic take on black hole entropy. \cite{Prunkl}   mention that black holes can feature in thermodynamic Carnot cycles just like any other thermodynamic system. Moreover, \cite{WallaceThermodynamics}   stresses that in all relevant respects black hole thermodynamics is not at all conceptually worse off than thermodynamics proper (granting of course that there is no empirical corroboration of black hole thermodynamics at the moment, nor likely to be had in the near future).  We do not want to repeat their specific arguments here, but we take them to   convincingly establish that black hole entropy is  thermodynamic entropy (in accordance with the consensus in the physics community).

With regard to  question (2), \cite{RovelliChirco} acknowledge that thermodynamic theories -- as far as we know -- have microstructure, but claim that the underlying degrees of freedom need not be novel or ``hidden''. Thermal radiation, for example, has an underlying microstructure which allegedly arises from quantization of the electromagnetic field, and is hence not novel.   \cite{RovelliChirco} state that, by analogy, the microstructure underlying black hole thermodynamics (or horizon thermodynamics) can arise from the quantization of the metric field alone.
We can defuse the sort of analogical reasoning at play here through usage of our terminology:   type I microstructure corresponds to no ``hidden degrees of freedom" in the language of \cite{RovelliChirco}, whereas type II microstructure corresponds to hidden degrees of freedom. Thermal radiation is underlied by electrodynamic fields, and they in turn can be seen to be underlied by quantum electrodynamic structure. Quantum electrodynamic structure is, thus, microstructure of type I with respect to classical electrodynamic structure, but type II microstructure with respect to thermal radiation. There is namely no sense in which electrodynamics can be seen to be  inspired from radiation thermodynamics. This contradicts the claim of \cite{RovelliChirco} that thermal radiation has type I microstructure (or no hidden degrees of freedom in their terminology). So their analogical reasoning cannot even take off.

However, one could argue from LQG (as \cite{RovelliChirco} do as well) or already from effective quantum field theory that black hole spacetimes (or the  gravitational metric structure   per se) have type I microstructure \textit{simpliciter}. Take the latter case: in an effective quantum field theory view on GR, the Bekenstein-Hawking formula for entropy is derivable from the partition function instantiated by the QFT path integral at play (see \cite{Gibbons:1976ue}). The calculation is formally in no way different from usual derivations of   entropy from the partition function in statistical mechanics (see for instance \cite{WallaceStatisticalMechanics} for a pedagogical introduction). But conceptually speaking, it must strike one as odd that a \textit{mere} formal analogy between QFT and statistical mechanics allows for calculating a meaningful statistical mechanical property (i.e. entropy) from a quantum field theoretic approach, \textit{without} there being any meaningful quantum field theoretical analogon to this property.  However, if one accepts the partition function formally obtained from a quantum field path integral as a true statistical mechanical object, one should   interpret the summation over different metric and matter fields in the partition function as the usual summation over all possible microstates. This would mean that the quantum structure obtained from quantizing the metric field of GR is sufficient microstructure to explain its thermodynamic behaviour.

This quantum field theory approach is, however, in conflict with the typical statistical mechanical derivation of thermodynamic entropy. The latter, namely, is based on  type~II microstructure, such as the positions of particles for a gas. If black hole entropy is a standard thermodynamic entropy then you might say that it also requires a standard statistical mechanical explanation in terms of type II microstructure. String theory indeed provides such a typical statistical mechanical derivation, because the Bekenstein-Hawking formula has been derived by counting certain string microstates or D-brane configurations (see \cite{Horowitz} for a review). In other words, black hole entropy is interpreted in string theory as    Boltzmann entropy, and hence is due to underlying type II microstructure. But, of course, we can only concur that it remains a physical possibility that black hole entropy   arises due to type I microstructure, such as entanglement (see also \cite{Bianchi:2012ev} for a discussion on this topic).

\subsection{Holographic principle}
\label{holography}

The holographic principle is generally viewed as an important guiding principle in the search for a theory of quantum gravity. It puts a precise limit on the number of fundamental degrees of freedom associated to spacetime regions, by stating that the maximal amount of information inside a given spacetime region cannot exceed the area of its boundary (measured in Planck units). 
In this section we are particularly interested in whether the holographic principle   entails anything about the nature of a putative microstructure of GR. Before we explain how the  principle might be viewed as a hint for type II microstructure -- and hence for emergent gravity -- we first give a more precise definition of it. 

The \emph{holographic principle}, originally proposed by \cite{tHooft:1993dmi} and followed up by \cite{Susskind:1994vu}, is based on the following simple argument. Consider a spatial spherically symmetric volume $V$ in an asymptotically flat spacetime that is bounded by a surface $B$. Now the  entropy and number of degrees of freedom inside $B$ are maximized if the volume contains a black hole. This is because, if one tries to increase the entropy by putting in more and more matter, at some point it will collapse into a black hole. As we saw in the previous section,  thermodynamical arguments suggest that  black holes carry an entropy proportional to their horizon area $A(B)$. If one were to throw more matter into the black hole, then its size would grow and it would not fit inside the surface $B$ any more. Therefore, by  dropping the specific assumption of spherical symmetry and asymptotic structure, 't Hooft  concluded that the number of degrees of freedom contained in any spatial region will not exceed the area of the region's boundary. 

Later \cite{Bousso}  gave the holographic principle a covariant formulation. Firstly, the holographic principle in its most general form does not hold for spacelike volumes, but   for  light sheets and their associated spacelike boundary. Secondly, a rendering of the holographic principle in terms of degrees of freedom already amounts to an interpretation of (black hole) entropy in a general-relativistic setting. Whether entropy in a gravitational context is due to a Boltzmannian-type counting procedure, due to entanglement across the surface or even a third option, is however up to debate (string theorists typically promote the first, LQG proponents favour the second one or even both at once, see \cite{Rovelli}).
   The interpretation-neutral version of the covariant entropy bound now reads as follows:

\begin{quote}
``The entropy on any light sheet of a surface $B$ will not exceed the area of $B$"\footnote{\cite{Bousso}, p. 19.}
$$S[L(B)] \leq \frac{k_B c^3}{\hbar}  \frac{A(B)}{4 G} \, . $$
\end{quote}

\noindent Here $L$ is a light sheet ``constructed by following light rays that emanate from the [spacelike] surface $B$, as long as they are not expanding". The covariant entropy bound can be shown to reduce to the original version of the holographic principle by 't Hooft and Susskind  for spatial volumes and their associated boundary surfaces.

The holographic principle has led to important progress in quantum gravity research. Already 't Hooft put forward a much stronger statement  than the entropy bound,  namely that   all the fundamental degrees of freedom inside a region can be described by a quantum mechanical theory living on the region's boundary (which is now known as a \emph{holographic duality}). Such holographic duality relations have indeed been found, most notably the AdS/CFT correspondence, and play a central role in quantum gravity research nowadays. We want to stress though that the holographic principle  (or, more precisely, the covariant entropy bound) should not be conflated with holographic duality relations. The latter is a conjecture about the equivalence of two, at first sight, unrelated theories, whereas the former is a bound on the gravitational entropy.  

Physicists often interpret the holographic relation as an emergent relation, in the sense that a theory of gravity in a given spacetime emerges from a quantum theory without gravity on the boundary. However, \cite{Dieks} and \cite{Haro} among others\footnote{See also the earlier work by \cite{Rickles} and \cite{Teh}.} worked out that emergence (also in our sense of underlying microstructure) and holographic duality are \textit{prima facie} independent notions. Only if the holographic duality is approximate at a certain level,   can it turn out that either the bulk or the boundary is more fundamental. In the latter case,  the bulk system can be seen to have underlying microstructure in terms of the (microscopic) boundary system. The judge is still out, though, on whether the AdS/CFT correspondence is an emergence relation or not, because it is still only proven in a large $N$ limit (for semi-classical bulk gravity).

We take it, however, that the holographic principle itself (and not just  the specific notion of holographic duality) already contains a hint for emergent gravity which may be formalized roughly as follows (cf. the introduction of \cite{Bousso} for a similar argument):

 \begin{enumerate}
	\item In a local QFT the total entropy of a spatial region scales
	with its volume.
	\item The holographic principle implies that the total entropy associated to a
	spatial region   does not scale with its volume, but with the area of the boundary of the region.
	\thus The     microtheory underlying gravity is
	non-local.
	\thus Non-local degrees of freedom are typically an
	example of type II microstructure.
\end{enumerate}

 \noindent Let us expand on this. We defined emergent gravity in section \ref{whatiseg} by means of type II microstructure. Now the holographic principle is not only concerned with the \emph{number} of fundamental degrees of freedom of quantum gravity, but it is also suggestive of the \emph{kind} of degrees of freedom that underly GR. As is well known, the holographic principle is in conflict with the conventional wisdom in statistical mechanics or quantum field theory that the number of degrees of freedom scales with the  volume size of a system. Conventional quantum field theories are interaction-local.\footnote{By interaction locality, we mean what is more usually dubbed `micro-causality', i.e. that spacelike separated operator field values do not commute. For a scalar field $\phi$, this means that $[\phi(x), \phi(y)]=0$ provided that $x$ and $y$ are spacelike separated points.} Since degrees of freedom are linked to each spatial point, the information content of a spatial region grows with the volume. The holographic principle, on the other hand, implies that the entropy is bounded by the area of spacelike surfaces. Hence, the holographic principle casts  doubt on the locality of the fundamental degrees of freedom of gravity, and this might mean that the underlying microstructure of GR is `non-trivial' (or of type II). 

Still, the non-locality of the underlying degrees of freedom does not need to imply that the microstructure is of type II. It is true that in  string theory, for example,  black hole entropy is accounted for by non-local fundamental objects such as strings and D-branes. However, LQG (which is arguably a form of quantum GR) does for instance feature interaction non-locality but is based on quantization of the standard gravitational degrees of freedom. So whether the non-locality of the fundamental degrees of freedom is a hint towards emergent gravity is not clear.

\section{Conclusion: From hints to arguments}
\label{conclusion}

We have assessed four different hints for the emergent nature of gravity: (1) universal coupling, (2) non-renormalizability, (3) black hole thermodynamics, and (4) the holographic principle. Whether you take any of these hints to be strong indications in favour of an emergent paradigm or not, depends to a good degree on (not fully uncontroversial) background assumptions. If you are committed to the geometric view,   universal coupling  is just a consequence of the fact that gravity relates to the   geometry of spacetime; whereas if you favour a physical view of the metric field, universal coupling \emph{is} suggestive of emergent gravity. Likewise, whether quantum GR is renormalizable or not, or whether a proper quantum gravity approach like LQG is able to cope with the holographic principle  are open issues. String theorists and quantum general relativity proponents   are likely to  take these issues to be settled, although their answers will differ. Whether black hole thermodynamics on its own is a hint, depends on whether   black hole entropy can be explained through quantized GR, or only by a type II microtheory such as string theory. And although Jacobson's reinterpretation of GR in terms of horizon thermodynamics provides a clear case of how gravity can be due to underlying type II microstructure, this reinterpretation might be contested. Having said this, we take it to have established why under acceptable background assumptions the hints make a case for emergent gravity.

However, we do not want to limit ourselves to presenting a critical account of individual suggestive features for emergent gravity, but rather strive forward by mounting exemplary analogical arguments based on these hints all together. For instance, the features of GR we referred to as possible hints in the previous sections can star in an argument from example:

\begin{quote}
\textbf{Argument from example}
	\item  Models of hydrodynamics have features $F_1, ..., F_n$.
	\item Whenever a model of hydrodynamics applies to a domain of the world, a model of a corresponding type II microtheory applies as well. In particular, there is type II microstructure.
	\item Models of thermodynamics have features $F_1, .., F_n$.
	\item Whenever a model of thermodynamics applies to a domain of the world, a model of a corresponding type II microtheory applies as well. In particular, there is type II microstructure.
	\item Models of GR have features $F_1, ..., F_n$.
	\thus Whenever a model of GR applies to a domain of the world, a model of a corresponding type II microtheory applies as well. In particular, there is type II microstructure.
\end{quote}

\noindent
$\{F_1, .., F_n\}$ is a set of features including (1) thermodynamic aspects and (2) universality.  Another form of analogical argument which can be used in favour of emergent gravity is

\begin{quote}
\textbf{Argument from likeness}
\item   Models of hydrodynamics and GR are similar with respect to features $\{F_1, .., F_n\}$.

\item   Models of hydrodynamics and GR are dissimilar with respect to features $\{G_1, .., G_m\}$.

\item The features $\{F_1, .., F_n\}$ weigh out the features $\{G_1, .., G_m\}$.

\item Whenever a model of hydrodynamics applies to a domain of the world, a model of a corresponding type II microtheory applies as well.  In particular, there is type II microstructure.
 
\thus The models of hydrodynamics and GR are decisively  similar. Hence, whenever a model of GR applies to a domain of the world, a model of a corresponding type II microtheory applies as well. In particular, there is type II microstructure. 
\end{quote}

\noindent
where the set $\{G_1, .., G_m\}$ could include, for instance, background-independence.\footnote{GR is background-independent -- see \cite{Belot} for an explication of this notion -- while (standard) hydrodynamics is not.}

So, rather than assessing the strength of individual features, one could argue that several features together make a stronger case than they would  on their own. To assess the strength of such holistic arguments, one standardly considers several criteria such as:  the number of similarities involved; the overall structural or syntactical similarity; common causes or general laws featuring in analogies (strong causal ties/law-like relations in the target domain and the source domain, respectively, are desired); usage of generalizations (as little as possible); and the material analogy (concrete, observable similarities between the source and the target domain are desirable).\footnote{See \cite{sep-reasoning-analogy}.} Applying these criteria to the argument of likeness above, the partial lack of material analogy for certain similarities between GR and hydrodynamics will strike some as problematic. Whereas, for instance, local thermodynamic features of hydrodynamic theories have been observed (and are observable in a straightforward sense), the same cannot be said about GR (black hole thermodynamics is still awaiting   \textit{direct} empirical tests.) 
On the other hand, in particular the mere amount of similarities between gravity and hydrodynamics, as well as their very nature  -- local thermodynamics and universality  are all law-like relations, albeit on a formal level -- speak in favour of the argument.

In our discussion of hints we left out two putative challenges standardly mentioned in the context of emergent gravity: (1) the Weinberg-Witten theorem,\footnote{ The (second)  theorem by \cite{Weinberg:1980kq} states that   theories with a conserved, Lorentz covariant stress-energy tensor cannot contain massless particles of spin $j>1$. This rules out the possibility that the graviton is an elementary or composite particle in an ordinary renormalizable quantum field theory. There are several interesting ways to evade this theorem, as discussed for instance by \cite{Loebbert:2008zz} and \cite{Carlip}.}  and (2) the difficulty of emergence of diffeomorphism invariant systems from non-diffeomorphism invariant ones. These challenges   have been sufficiently dealt with -- or rather dismissed -- elsewhere in the literature (see \cite{Carlip} for a dismissal of the Weinberg-Witten theorem as a challenge to emergent gravity). That diffeomorphism invariant systems can emerge out of non-diffeomorphism invariant systems, has been demonstrated both by the spin-2 account and analogue gravity models as a proof of principle.

Concluding, we have taken up the challenge of making explicit arguments in favour of an emergent gravity paradigm which are referred to in the literature or simply roam around in the community. Needless to say, the depiction of the arguments at play cannot give sufficient credit to all possible variants. 
Let us anticipate another possible objection, namely that it is unfair to appraise arguments which are only meant as mere heuristics or intuitions in the first place. However, we take this point to be ill-founded. That arguments can only be plausibility arguments at the heuristic level, does not mean that they are immune to scrutiny and critical assessment tout court. The philosopher's of physics job in the process of discovery of quantum gravity -- so we believe -- should amount to providing exactly this kind of assessments. The current paper is our contribution to this effort.

\subsection*{Acknowledgments}

We are grateful to   Harvey Brown, Karen Crowther, James Read, Jeroen van Dongen, Sebastian de Haro, Niels Martens, Kian Salimkhani, Erik Verlinde,   Christian W\"{u}thrich and two anonymous referees for useful comments on the paper and/or general discussions on emergent gravity. 
We would also like to thank the History and 
Philosophy of Gravity group at the University of Amsterdam,
and the audiences of the EPSA conference in Exeter and the  ``Thinking about Space and Time''  conference in Bern for useful feedback (both held in  summer 2017). The work of M.V. was supported by the Spinoza Grant and by the Delta Institute for Theoretical Physics, both financed by the Dutch Science Organisation (NWO).   N.L. is thankful for financial support from the Swiss Science Foundations (SNF) (project number $105212\_165702$).

 \bibliography{references}

\begin{thebibliography}{86}
\providecommand{\natexlab}[1]{#1}
\providecommand{\url}[1]{\texttt{#1}}
\expandafter\ifx\csname urlstyle\endcsname\relax
  \providecommand{\doi}[1]{doi: #1}\else
  \providecommand{\doi}{doi: \begingroup \urlstyle{rm}\Url}\fi

\bibitem[Aharony and Banks(1999)]{Banks}
Ofer Aharony and Tom Banks.
\newblock {Note on the quantum mechanics of M theory}.
\newblock \emph{{Journal of High Energy Physics}}, 03:\penalty0 016, 1999.

\bibitem[Aharony et~al.(2000)Aharony, Gubser, Maldacena, Ooguri, and
  Oz]{Aharony}
Ofer Aharony, Steven~S. Gubser, Juan~M. Maldacena, Hirosi Ooguri, and Yaron Oz.
\newblock {Large N field theories, string theory and gravity}.
\newblock \emph{Phys. Rept.}, 323:\penalty0 183--386, 2000.

\bibitem[Bain(2013{\natexlab{a}})]{Bain}
Jonathan Bain.
\newblock {The emergence of spacetime in condensed matter approaches to quantum
  gravity}.
\newblock \emph{Studies in History and Philosophy of Modern Physics},
  44\penalty0 (3):\penalty0 338--345, 2013{\natexlab{a}}.

\bibitem[Bain(2013{\natexlab{b}})]{Bain2013}
Jonathan Bain.
\newblock Effective field theories.
\newblock In B.~Batterman, editor, \emph{The Oxford Handbook of Philosophy of
  Physics}, pages 224--254. Oxford University Press, New York,
  2013{\natexlab{b}}.

\bibitem[Banks et~al.(1997)Banks, Fischler, Shenker, and Susskind]{Banks1997}
Tom Banks, Willy Fischler, Stephen~H. Shenker, and Leonard Susskind.
\newblock {M theory as a matrix model: A Conjecture}.
\newblock \emph{{Physical Review D}}, 55:\penalty0 5112--5128, 1997.

\bibitem[Barcelo et~al.(2005)Barcelo, Liberati, and Visser]{AnalogueGravity}
Carlos Barcelo, Stefano Liberati, and Matt Visser.
\newblock {Analogue gravity}.
\newblock \emph{{Living Reviews in Relativity}}, 8:\penalty0 12, 2005.

\bibitem[Bardeen et~al.(1973)Bardeen, Carter, and
  Hawking]{BardeenCarterHawking}
James~M. Bardeen, Brandon Carter, and Stephen~W. Hawking.
\newblock The four laws of black hole mechanics.
\newblock \emph{Comm. Math. Phys.}, 31\penalty0 (2):\penalty0 161--170, 1973.

\bibitem[Bartha(2016)]{sep-reasoning-analogy}
Paul Bartha.
\newblock {Analogy and Analogical Reasoning}.
\newblock In E.~N. Zalta, editor, \emph{The Stanford Encyclopedia of
  Philosophy}. Metaphysics Research Lab, Stanford University, winter 2016
  edition, 2016.

\bibitem[Becker et~al.(2006)Becker, Becker, and Schwarz]{Becker:2007zj}
K.~Becker, M.~Becker, and J.~H. Schwarz.
\newblock \emph{{String theory and M-theory: A modern introduction}}.
\newblock Cambridge University Press, 2006.

\bibitem[Bekenstein(1972)]{BekensteinOriginal}
Jacob~D. Bekenstein.
\newblock {Black holes and the second law}.
\newblock \emph{Lettere Al Nuovo Cimento (1971--1985)}, 4\penalty0
  (15):\penalty0 737--740, 1972.

\bibitem[Belot(2011)]{Belot}
Gordon Belot.
\newblock Background-independence.
\newblock \emph{General Relativity and Gravitation}, 43\penalty0 (10):\penalty0
  2865--2884, 2011.

\bibitem[Berenstein(2006)]{Berenstein2005}
David Berenstein.
\newblock {Large N BPS states and emergent quantum gravity}.
\newblock \emph{{Journal of High Energy Physics}}, 01:\penalty0 125, 2006.

\bibitem[Bianchi and Myers(2014)]{Bianchi:2012ev}
Eugenio Bianchi and Robert~C. Myers.
\newblock {On the architecture of spacetime geometry}.
\newblock \emph{{Classical and Quantum Gravity}}, 31:\penalty0 214002, 2014.

\bibitem[Bousso(2002)]{Bousso}
Raphael Bousso.
\newblock The holographic principle.
\newblock \emph{Reviews of Modern Physics}, 74\penalty0 (3):\penalty0 825,
  2002.

\bibitem[Brown(2005)]{Brown}
Harvey~R. Brown.
\newblock \emph{Physical relativity: Space-time structure from a dynamical
  perspective}.
\newblock Oxford University Press, 2005.

\bibitem[Brown and Read(2016)]{BrownRead}
Harvey~R. Brown and James Read.
\newblock Clarifying possible misconceptions in the foundations of general
  relativity.
\newblock \emph{American Journal of Physics}, 84\penalty0 (5):\penalty0
  327--334, 2016.

\bibitem[Burgess(2004)]{Burgess2003}
Clifford~P. Burgess.
\newblock {Quantum gravity in everyday life: General relativity as an effective
  field theory}.
\newblock \emph{{Living Reviews in Relativity}}, 7:\penalty0 5--56, 2004.

\bibitem[Butterfield(2011)]{ButterfieldEmergence}
Jeremy Butterfield.
\newblock Emergence, reduction and supervenience: a varied landscape.
\newblock \emph{Foundations of Physics}, 41\penalty0 (6):\penalty0 920--959,
  2011.

\bibitem[Cao(1998)]{CaoConceptualDevelopments}
Tian~Yu Cao.
\newblock \emph{Conceptual developments of 20th century field theories}.
\newblock Cambridge University Press, 1998.

\bibitem[Cao and Schweber(1993)]{CaoSchweber}
Tian~Yu Cao and Silvan~S. Schweber.
\newblock The conceptual foundations and the philosophical aspects of
  renormalization theory.
\newblock \emph{Synthese}, 97\penalty0 (1):\penalty0 33--108, 1993.

\bibitem[Carlip(2014)]{Carlip}
Steven Carlip.
\newblock Challenges for emergent gravity.
\newblock \emph{Studies in History and Philosophy of Modern Physics},
  46:\penalty0 200--208, 2014.

\bibitem[Chirco et~al.(2014)Chirco, Haggard, Riello, and
  Rovelli]{RovelliChirco}
Goffredo Chirco, Hal~M. Haggard, Aldo Riello, and Carlo Rovelli.
\newblock Spacetime thermodynamics without hidden degrees of freedom.
\newblock \emph{Physical Review D}, 90\penalty0 (4):\penalty0 044044, 2014.

\bibitem[Crowther(2013)]{Crowther}
Karen Crowther.
\newblock {Emergent spacetime according to effective field theory: From
  top-down and bottom-up}.
\newblock \emph{Studies in History and Philosophy of Modern Physics},
  44\penalty0 (3):\penalty0 321--328, 2013.

\bibitem[Crowther(2016)]{Crowther2016}
Karen Crowther.
\newblock \emph{{Effective Spacetime: Understanding Emergence in Effective
  Field Theory and Quantum Gravity}}.
\newblock Springer, Heidelberg, 2016.

\bibitem[Curiel(2017)]{Curiel}
Erik Curiel.
\newblock {A Primer on Energy Conditions}.
\newblock In G.~Schiemann D.~Lehmkuhl and E.~Scholz, editors, \emph{{Towards a
  Theory of Spacetime Theories}}, volume~13 of \emph{{Einstein Studies}}, pages
  43--104. Birkhauser, 2017.

\bibitem[de~Haro(2017{\natexlab{a}})]{Haro}
Sebastian de~Haro.
\newblock {Dualities and emergent gravity: Gauge/gravity duality}.
\newblock \emph{Studies in History and Philosophy of Modern Physics},
  B59:\penalty0 109--125, 2017{\natexlab{a}}.

\bibitem[de~Haro(2017{\natexlab{b}})]{deHaronew}
Sebastian de~Haro.
\newblock {Towards a Theory of Emergence for the Physical Sciences}.
\newblock \emph{Unpublished draft}, 2017{\natexlab{b}}.

\bibitem[Deser(1970)]{Deser:1969wk}
Stanley Deser.
\newblock {Selfinteraction and gauge invariance}.
\newblock \emph{Gen. Rel. Grav.}, 1:\penalty0 9--18, 1970.

\bibitem[Dieks et~al.(2015)Dieks, van Dongen, and de~Haro]{Dieks}
Dennis Dieks, Jeroen van Dongen, and Sebastian de~Haro.
\newblock {Emergence in holographic scenarios for gravity}.
\newblock \emph{Studies in History and Philosophy of Modern Physics},
  B52:\penalty0 203--216, 2015.

\bibitem[Doboszewski and Linnemann(2017)]{DoboszewskiLinnemann}
Juliusz Doboszewski and Niels Linnemann.
\newblock How not to establish the non-renormalizability of gravity.
\newblock \emph{Foundations of Physics}, pages 1--16, 2017.

\bibitem[Donoghue(2012)]{Donoghue}
John~F. Donoghue.
\newblock {The effective field theory treatment of quantum gravity}.
\newblock \emph{AIP Conf. Proc.}, 1483:\penalty0 73--94, 2012.

\bibitem[Dougherty and Callender(2016)]{DoughertyCallender}
John Dougherty and Craig Callender.
\newblock {Black Hole Thermodynamics: More Than an Analogy?}
\newblock \emph{Preprint}, 2016.

\bibitem[Earman and Norton(1987)]{NortonEarman}
John Earman and John Norton.
\newblock {What Price Spacetime Substantivalism? The Hole Story}.
\newblock \emph{The British Journal for the Philosophy of Science}, 38\penalty0
  (4):\penalty0 515--525, 1987.

\bibitem[Feynman(2003)]{Feynman}
Richard~P. Feynman.
\newblock \emph{Feynman lectures on gravitation}, 2003.

\bibitem[Friedman(2014)]{Friedman}
Michael Friedman.
\newblock \emph{Foundations of space-time theories: Relativistic physics and
  philosophy of science}.
\newblock Princeton University Press, 2014.

\bibitem[Gibbons and Hawking(1977)]{Gibbons:1976ue}
Gary~W. Gibbons and Stephen~W. Hawking.
\newblock {Action Integrals and Partition Functions in Quantum Gravity}.
\newblock \emph{{Physical Review D}}, 15:\penalty0 2752--2756, 1977.

\bibitem[Goroff and Sagnotti(1986)]{GoroffSagnotti}
Marc~H. Goroff and Augusto Sagnotti.
\newblock {The ultraviolet behavior of Einstein gravity}.
\newblock \emph{Nuclear Physics B}, 266\penalty0 (3):\penalty0 709--736, 1986.

\bibitem[Hawking(1975)]{Hawking}
Stephen~W. Hawking.
\newblock Particle creation by black holes.
\newblock \emph{Communications in mathematical physics}, 43\penalty0
  (3):\penalty0 199--220, 1975.

\bibitem[Horowitz and Polchinski(2006)]{Horowitz}
Gary~T. Horowitz and Joseph Polchinski.
\newblock Gauge/gravity duality.
\newblock In D.~Oriti, editor, \emph{Approaches to quantum gravity: Toward a
  New Understanding of Space, Time and Matter}, pages 169--186. Cambridge
  University Press, 2006.

\bibitem[Hu(2009)]{Hu}
Bei~Lok Hu.
\newblock {Emergent/quantum gravity: macro/micro structures of spacetime}.
\newblock \emph{J. Phys. Conf. Ser.}, 174:\penalty0 012015, 2009.

\bibitem[Huggett and W{\"u}thrich(2013)]{HuggettWuethrich}
Nick Huggett and Christian W{\"u}thrich.
\newblock Emergent spacetime and empirical (in) coherence.
\newblock \emph{Studies in History and Philosophy of Modern Physics},
  44\penalty0 (3):\penalty0 276--285, 2013.

\bibitem[Jacobson(1995)]{Jacobson}
Ted Jacobson.
\newblock {Thermodynamics of Spacetime: the Einstein Equation of State}.
\newblock \emph{Physical Review Letters}, 75\penalty0 (7):\penalty0 1260, 1995.

\bibitem[Jacobson(2016)]{Jacobson2015}
Ted Jacobson.
\newblock {Entanglement Equilibrium and the Einstein Equation}.
\newblock \emph{{Physical Review Letters}}, 116\penalty0 (20):\penalty0 201101,
  2016.

\bibitem[Jacobson and Parentani(2003)]{JacobsonHorizon}
Ted Jacobson and Renaud Parentani.
\newblock Horizon entropy.
\newblock \emph{Foundations of Physics}, 33\penalty0 (2):\penalty0 323--348,
  2003.

\bibitem[Kiefer(2007)]{KieferTextbook}
Claus Kiefer.
\newblock \emph{Quantum Gravity}, volume 136.
\newblock Oxford University Press, 2 edition, 2007.

\bibitem[Knox(2013)]{Knox2013}
Eleanor Knox.
\newblock {Effective spacetime geometry}.
\newblock \emph{Studies in History and Philosophy of Modern Physics},
  44\penalty0 (3):\penalty0 346--356, 2013.

\bibitem[Lam and W{\"u}thrich(2017)]{LamWuthrich}
Vincent Lam and Christian W{\"u}thrich.
\newblock Spacetime is as spacetime does.
\newblock \emph{Unpublished Draft}, 2017.

\bibitem[Loebbert(2008)]{Loebbert:2008zz}
Florian Loebbert.
\newblock {The Weinberg-Witten theorem on massless particles: An Essay}.
\newblock \emph{Annalen Phys.}, 17:\penalty0 803--829, 2008.

\bibitem[Maldacena(1999)]{Maldacena}
Juan~M. Maldacena.
\newblock {The Large N limit of superconformal field theories and
  supergravity}.
\newblock \emph{Int. J. Theor. Phys.}, 38:\penalty0 1113--1133, 1999.
\newblock [Adv. Theor. Math. Phys. 2, 231 (1998)].

\bibitem[Mattingly(2014)]{Mattingly}
James Mattingly.
\newblock Unprincipled microgravity.
\newblock \emph{Studies in History and Philosophy of Modern Physics},
  46:\penalty0 179--185, 2014.

\bibitem[Maudlin(2012)]{Maudlin}
Tim Maudlin.
\newblock \emph{Philosophy of physics: Space and time}.
\newblock Princeton University Press, 2012.

\bibitem[Misner et~al.(1973)Misner, Thorne, and Wheeler]{Gravitation}
Charles~W. Misner, Kip~S. Thorne, and John~A. Wheeler.
\newblock \emph{Gravitation}.
\newblock Macmillan, 1973.

\bibitem[Nickles(1973)]{Nickles}
Thomas Nickles.
\newblock Two concepts of intertheoretic reduction.
\newblock \emph{The Journal of Philosophy}, 70\penalty0 (7):\penalty0 181--201,
  1973.

\bibitem[Padmanabhan(2010)]{Padmanabhan1}
Thanu Padmanabhan.
\newblock {Thermodynamical Aspects of Gravity: New insights}.
\newblock \emph{Rept. Prog. Phys.}, 73:\penalty0 046901, 2010.

\bibitem[Padmanabhan(2016)]{Padmanabhan2}
Thanu Padmanabhan.
\newblock {Exploring the Nature of Gravity}.
\newblock In \emph{{The Planck Scale II Wroclaw, Poland, September 7-12,
  2015}}, 2016.

\bibitem[Parisi(1975)]{Parisi}
Giorgio Parisi.
\newblock {The theory of non-renormalizable interactions: The large N
  expansion}.
\newblock \emph{Nuclear Physics B}, 100\penalty0 (2):\penalty0 368--388, 1975.

\bibitem[Peskin and Schroeder(1995)]{PeskinSchroeder}
Michael~E. Peskin and Daniel~V. Schroeder.
\newblock \emph{{An Introduction to quantum field theory}}.
\newblock Addison-Wesley, Reading, USA, 1995.

\bibitem[Polchinski(2007{\natexlab{a}})]{Polchinski:1998rq}
J.~Polchinski.
\newblock \emph{{String theory. Vol. 1: An introduction to the bosonic
  string}}.
\newblock Cambridge University Press, 2007{\natexlab{a}}.

\bibitem[Polchinski(2007{\natexlab{b}})]{Polchinski:1998rr}
J.~Polchinski.
\newblock \emph{{String theory. Vol. 2: Superstring theory and beyond}}.
\newblock Cambridge University Press, 2007{\natexlab{b}}.

\bibitem[Prunkl and Timpson(2017)]{Prunkl}
Carina Prunkl and Christopher Timpson.
\newblock {Black Hole Entropy is Entropy and not (necessarily) Information}.
\newblock \emph{Unpublished draft}, 2017.

\bibitem[Read et~al.(2017)Read, Brown, and Lehmkuhl]{ReadBrownLehmkuhl}
James Read, Harvey~R. Brown, and Dennis Lehmkuhl.
\newblock Two miracles of general relativity.
\newblock \emph{Unpublished draft}, 2017.

\bibitem[Rickles(2013)]{Rickles}
Dean Rickles.
\newblock {AdS/CFT duality and the emergence of spacetime}.
\newblock \emph{Studies in the History and Philosophy of Modern Physics},
  B44:\penalty0 312--320, 2013.

\bibitem[Rovelli(2004)]{Rovelli}
Carlo Rovelli.
\newblock \emph{{Quantum gravity}}.
\newblock Cambridge Monographs on Mathematical Physics. Cambridge University
  Press, 2004.

\bibitem[Sakharov(1968)]{Sakharov1967}
Andrei~D. Sakharov.
\newblock {Vacuum quantum fluctuations in curved space and the theory of
  gravitation}.
\newblock \emph{Sov. Phys. Dokl.}, 12:\penalty0 1040--1041, 1968.
\newblock [Gen. Rel. Grav. 32, 365 (2000)].

\bibitem[Seiberg(2006)]{Seiberg}
Nathan Seiberg.
\newblock {Emergent spacetime}.
\newblock In \emph{{The Quantum Structure of Space and Time: Proceedings of the
  23rd Solvay Conference on Physics. Brussels, Belgium. 1 - 3 December 2005}},
  pages 163--178, 2006.

\bibitem[Shomer(2008)]{Shomer}
Assaf Shomer.
\newblock {A Pedagogical explanation for the non-renormalizability of gravity.}
\newblock \emph{arXiv: 0709.3555 [hep-th]}, 2008.

\bibitem[Sindoni(2012)]{Sindoni2012}
Lorenzo Sindoni.
\newblock {Emergent models for gravity: An overview of microscopic models}.
\newblock \emph{Symmetry, Integrability and Geometry: Methods and
  Applications}, 8, 2012.

\bibitem[Susskind(1995)]{Susskind:1994vu}
Leonard Susskind.
\newblock {The world as a hologram}.
\newblock \emph{{Journal of Mathematical Physics}}, 36:\penalty0 6377--6396,
  1995.

\bibitem['t~Hooft(1993)]{tHooft:1993dmi}
Gerard 't~Hooft.
\newblock {Dimensional reduction in quantum gravity}.
\newblock In \emph{{Salamfest 1993:0284-296}}, pages 0284--296, 1993.

\bibitem[Teh(2013)]{Teh}
Nicholas~J. Teh.
\newblock {Holography and Emergence}.
\newblock \emph{Studies in History and Philosophy of Modern Physics},
  44\penalty0 (3):\penalty0 300--311, 2013.

\bibitem[Thorne et~al.(1973)Thorne, Lee, and Lightman]{ThorneLee}
Kip~S. Thorne, David~L. Lee, and Alan~P. Lightman.
\newblock Foundations for a theory of gravitation theories.
\newblock \emph{Physical Review D}, 7\penalty0 (12):\penalty0 3563, 1973.

\bibitem[Unruh(1995)]{AnalogueUnruh}
William~G. Unruh.
\newblock Sonic analogue of black holes and the effects of high frequencies on
  black hole evaporation.
\newblock \emph{Physical Review D}, 51:\penalty0 2827--2838, Mar 1995.

\bibitem[Verlinde(2011)]{Verlinde}
Erik~P. Verlinde.
\newblock {On the origin of gravity and the laws of Newton}.
\newblock \emph{Journal of High Energy Physics}, 2011\penalty0 (4):\penalty0
  1--27, 2011.

\bibitem[Verlinde(2017)]{Verlinde2016}
Erik~P. Verlinde.
\newblock {Emergent gravity and the dark universe}.
\newblock \emph{SciPost Phys.}, 2\penalty0 (3):\penalty0 016, 2017.

\bibitem[Visser(2002)]{Visser2002}
Matt Visser.
\newblock {Sakharov's induced gravity: A Modern perspective}.
\newblock \emph{Mod. Phys. Lett.}, A17:\penalty0 977--992, 2002.

\bibitem[Vistarini(2017)]{Vistarini}
Tiziana Vistarini.
\newblock Holographic space and time: Emergent in what sense?
\newblock \emph{Studies in History and Philosophy of Science Part B: Studies in
  History and Philosophy of Modern Physics}, 59:\penalty0 126--135, 2017.

\bibitem[Wald(2001)]{Wald2001}
Robert~M. Wald.
\newblock {The thermodynamics of black holes}.
\newblock \emph{{Living Reviews in Relativity}}, 4:\penalty0 6, 2001.

\bibitem[Wallace(2017{\natexlab{a}})]{WallaceStatisticalMechanics}
David Wallace.
\newblock The case for black hole thermodynamics, part ii: statistical
  mechanics.
\newblock \emph{arXiv:1710.02725 [gr-qc]}, 2017{\natexlab{a}}.

\bibitem[Wallace(2017{\natexlab{b}})]{WallaceThermodynamics}
David Wallace.
\newblock The case for black hole thermodynamics, part i: phenomenological
  thermodynamics.
\newblock \emph{arXiv:1710.02724 [gr-qc]}, 2017{\natexlab{b}}.

\bibitem[Weinberg(1979)]{Weinberg}
Steven Weinberg.
\newblock Ultraviolet divergencies in quantum theories of gravitation.
\newblock In S.~Hawking and W.~Israel, editors, \emph{General relativity, an
  Einstein Centenary survey}, pages 790--831. Cambridge University Press, 1979.

\bibitem[Weinberg and Witten(1980)]{Weinberg:1980kq}
Steven Weinberg and Edward Witten.
\newblock {Limits on Massless Particles}.
\newblock \emph{Phys. Lett.}, 96B:\penalty0 59--62, 1980.

\bibitem[Will(2006)]{Clifford}
Clifford~M. Will.
\newblock The confrontation between general relativity and experiment.
\newblock \emph{Living Reviews in Relativity}, 9\penalty0 (1):\penalty0 3,
  2006.

\bibitem[Witten(1996)]{WittenReflections}
Edward Witten.
\newblock {Reflections on the fate of space-time}.
\newblock \emph{Physics Today}, 49N4:\penalty0 24--30, 1996.

\bibitem[W\"{u}thrich()]{WuthrichInformation}
Christian W\"{u}thrich.
\newblock Are black holes about information?
\newblock Forthcoming in R. Dawid, R. Dardashti, and K. Th{\'e}bault, editors,
  \emph{Epistemology of Fundamental Physics}, Cambridge University Press, 2017.

\bibitem[Zee(2010)]{Zee}
Anthony Zee.
\newblock \emph{Quantum field theory in a nutshell}.
\newblock Princeton University Press, 2010.

\bibitem[Zwiebach(2006)]{Zwiebach:2004tj}
B.~Zwiebach.
\newblock \emph{{A first course in string theory}}.
\newblock Cambridge University Press, 2006.

\end{thebibliography}

\end{document}